\documentclass[aps,prl,twocolumn,amssymb,superscriptaddress,showpacs]{revtex4-1}
\usepackage{float}
\usepackage{xcolor}
\usepackage{graphicx}
\usepackage{dcolumn}
\usepackage{bm}
\usepackage{hyperref}
\usepackage{color}
\usepackage{amsmath}
\usepackage{amsfonts}

\begin{document}

\title{Berry Curvature Engineering by Gating Two-Dimensional Antiferromagnets}

\author{Shiqiao Du}
\affiliation{State Key Laboratory of Low-Dimensional Quantum Physics, Department of Physics, Tsinghua University, Beijing 100084, China,}
\author{Peizhe Tang}
\email{E-mail: peizhe.tang@mpsd.mpg.de}
\affiliation{Max Planck Institute for the Structure and Dynamics of Matter, Center for Free-Electron Laser Science, Luruper Chaussee 149, 22761 Hamburg, Germany.}

\author{Jiaheng Li}
\affiliation{State Key Laboratory of Low-Dimensional Quantum Physics, Department of Physics, Tsinghua University, Beijing 100084, China,}
\author{Zuzhang Lin}
\affiliation{Institute for Advanced Study, Tsinghua University, Beijing 100084, China,}
\author{Yong Xu}
\email{E-mail: yongxu@mail.tsinghua.edu.cn}
\affiliation{State Key Laboratory of Low-Dimensional Quantum Physics, Department of Physics, Tsinghua University, Beijing 100084, China,}
\affiliation{Collaborative Innovation Center of Quantum Matter, Beijing 100084, China,}
\affiliation{RIKEN Center for Emergent Matter Science (CEMS), Wako, Saitama 351-0198, Japan,}

\author{Wenhui Duan}
\email{E-mail: dwh@phys.tsinghua.edu.cn}
\affiliation{State Key Laboratory of Low-Dimensional Quantum Physics, Department of Physics, Tsinghua University, Beijing 100084, China,}
\affiliation{Collaborative Innovation Center of Quantum Matter, Beijing 100084, China,}
\affiliation{Institute for Advanced Study, Tsinghua University, Beijing 100084, China,}

\author{Angel Rubio}
\email{E-mail: angel.rubio@mpsd.mpg.de}
\affiliation{Max Planck Institute for the Structure and Dynamics of Matter, Center for Free-Electron Laser Science, Luruper Chaussee 149, 22761 Hamburg, Germany,}
\affiliation{Nano-Bio Spectroscopy Group and ETSF, Dpto. Fisica de Materiales, Universidad del Pa\'{i}s Vasco UPV/EHU, 20018 San Sebasti\'{a}n, Spain}
\affiliation{Center for Computational Quantum Physics, Flatiron Institute, 162 Fifth Avenue, New York, NY 10010, USA.}

\begin{abstract}
Recent advances in tuning electronic, magnetic, and topological properties of two-dimensional (2D) magnets have opened a new frontier in the study of quantum physics and promised exciting possibilities for future quantum technologies. In this study, we find that the dual gate technology can well tune the electronic and topological properties of antiferromagnetic (AFM) even septuple-layer (SL) MnBi$_2$Te$_4$ thin films. Under an out-of-plane electric field that breaks $\mathcal{PT}$ symmetry, the Berry curvature of the thin film could be engineered efficiently, resulting in a huge change of anomalous Hall (AH) signal. Beyond the critical electric field, the double-SL MnBi$_2$Te$_4$ thin film becomes a Chern insulator with a high Chern number of 3. We further demonstrate that such 2D material can be used as an AFM switch via electric-field control of the AH signal. These discoveries inspire the design of low-power memory prototype for future AFM spintronic applications.
\end{abstract}

\date{ \today}
\maketitle


AFM spintronics, aiming to use antiferromagnets to complement or replace ferromagnets as active components of spintronic devices, opens a new era in the field of spintronics owing to its numerous advantages, including the robustness against perturbation of external magnetic field, the absence of stray field, and the ultrafast dynamics \cite{Baltz2018,Gomonay2017,jungwirth2016Anti,macdonald2011antiferromagnetic}. The key issue to design AFM spintronic devices is to find an efficient approach to manipulate and detect the magnetic or electronic quantum states of an antiferromagent. By using spin-transfer torque and spin-orbit torque \cite{Baltz2018}, external electric currents are able to write information by controlling the AFM order inside spintronic devices \cite{Zelezny2014,ZhouXF2018,marti2014room,ChenXZ2018,wang2018field} or by generating spin currents at interfaces between antiferromagnets and non-magnets \cite{Moriyama2018}. For example, via applying local current in the tetragonal CuMnAs thin film, AFM spin orientations can be manipulated for information storage, which can be readout via detecting the anisotropic magnetoresistance \cite{wadley2016electrical,bodnar2018writing}. Based on such proposal, room-temperature AFM memory cells have been fabricated experimentally \cite{olejnik2017}.

The electric detection of anomalous Hall effect (AHE) provides an alternative powerful scenario to design spintronic devices based on antiferromagnets. The physical origin of the AHE has been under debate for decades \cite{Nagaosa2010}. While, in recent years, people realize that Berry curvature $\mathbf{\Omega}(\textbf{k})$ in the momentum space plays an important role in generating the intrinsic AHE. Based on the symmetry argument, the Berry curvature could be non-zero in a system without the combination of time reversal symmetry ($\mathcal{T}$) and inversion symmetry ($\mathcal{P}$), namely $\mathcal{PT}$ symmetry. Such a constrain indicate the antiferromagnets with non-collinear magnetism and antiferromagnet heterostructures as potential material candidates, including Mn$_3$Ge \cite{Kiyohara2016,LiuJP2017,nayak2016large}, Mn$_3$Sn \cite{nakatsuji2015large}, Mn$_3$Ir\cite{ChenHua2014}, Mn$_3$Pt/BaTiO$_3$ \cite{liu2018electrical} and CrSb/Cr-doped (Bi,Sb)$_2$Te$_3$ superlattices \cite{he2017tailoring}. However, the external control of their AHE is challenging in practice. For AFM structures with collinear magnetism, the possible existence of $\mathcal{PT}$ symmetry guarantees the vanishing of Berry curvature, although non-trivial topological fermions may survive in some three-dimensional (3D) antiferromagnets \cite{tang2016dirac}. Therefore, to find an effective way to manipulate and engineer Berry curvature is essential to use AHE as the detection signal in AFM spintronic devices.


The recently discovered 3D AFM topological insulator (TI) MnBi$_2$Te$_4$ provides us a new chance to design topological quantum devices controlled by external fields. Similar to the prototype of 3D TI (\emph{e.g.}, Bi$_2$Se$_3$ family), MnBi$_2$Te$_4$ crystal is a layered material \cite{gong2018experimental,otrokov2018prediction,zhang2018topological,liJH2018}. There exist strong chemical bonding within each SL (consisting of Te-Bi-Te-Mn-Te-Bi-Te) and weak van der Waals bonding between SLs. Each Mn atom has a magnetic moment of $\sim$ 5 $\mu_{\mathbf{B}}$ ($\mu_{\mathbf{B}}$ is the Bohr magneton) and its spin orientation is along the out-of-plane $z$ direction \cite{OtrokovPRL2019,liJH2018}. In the ground state, a long-range ferromagnetic order is formed in each SL, and adjacent SLs couple with each other antiferromagnetically, displaying an N\'eel temperature $T_N$ $\sim$ 25 K \cite{zeugner2018chemical,CuiJH2019}. In contrast to AFM transition-metal oxides \cite{DongXY2016}, ultrathin films of layered MnBi$_2$Te$_4$ can be easily cleaved by mechanical exfoliation \cite{deng2019magnetic,liu2019quantum,ge2019} or grown via molecular beam epitaxy \cite{gong2018experimental}. The quantized Hall effect without Landau levels has been theoretically predicted and experimentally observed in these thin films \cite{liJH2018,OtrokovPRL2019,deng2019magnetic,liu2019quantum,ge2019}.

Herein, we focus on 2D AFM MnBi$_2$Te$_4$ thin film with double-SL whose intrinsic electronic property is topologically trivial. By using \emph{ab initio} calculations, we demonstrate that the out-of-plane electric field induced by dual-gate technology breaks $\mathcal{PT}$ symmetry in AFM double-SL, engineers its band structures significantly, and generates huge AH signals. Beyond the critical value, the electric field induces a topological phase transition, driving the double-SL MnBi$_2$Te$_4$ to be a quantized anomalous Hall (QAH) insulator with a high Chern number of 3. Based on these useful properties, we propose a prototype device as the electric field controlled topological AFM memory, whose performance is expected to be much higher than the current AFM random-access memory. Thanks to AFM ground state, its electric switch should be stable and robust under external magnetic field and the switching time is very fast.

The first-principles DFT calculations are performed by using the projector-augmented wave method implemented in Vienna \emph{ab initio} Simulation Package (VASP) with the GGA-PBE exchange-correlation functional \cite{KRESSE199615}. The energy cut-off for the plane wave basis is set to be 350 eV and the energy convergence criterion is $10^{-6}$ eV for self-consistent electronic-structure calculations. A 20 ${\rm \AA}$ vacuum layer is chosen to eliminate the periodic effect along the z-axis direction. The van der Waals interaction is included through the DFT-D3 method \cite{vdW-D3}, which has been tested to have the best performance on describing the lattice parameters compared with experiments. The crystal structures are relaxed until the force on each atom is less than 0.01 eV/${\rm \AA}$. A 24$\times$24$\times$1 $\Gamma$-centered $\emph{k}$-point mesh is sampled uniformly over the BZ. The GGA+$U$ method with $U$=4 eV is applied to describe the localized $3d$ orbitals of Mn atom in the electronic structure calculations. The electric field is applied along the non-periodic direction with the dipole corrections. The HSE06 hybrid functional is also applied to check the band gap value \cite{HSE06}. The maximally-localized Wannier functions are obtained by the Wannier90 packages \cite{wannier90}. The DFT calculated Bloch functions are projected onto Mn $d$, Te $p$ and Bi $s$ and $p$ orbitals to construct the corresponding Wannier functions. The edge-state and AH conductance is calculated based on the tight binding model from maximally localized Wannier functions \cite{WU2017}. To include the temperature effect, the Fermi-Dirac distribution function is applied in the Kubo-Greenwood formula for the calculation of AH conductance.


\begin{figure}
\includegraphics[width=1.0\columnwidth]{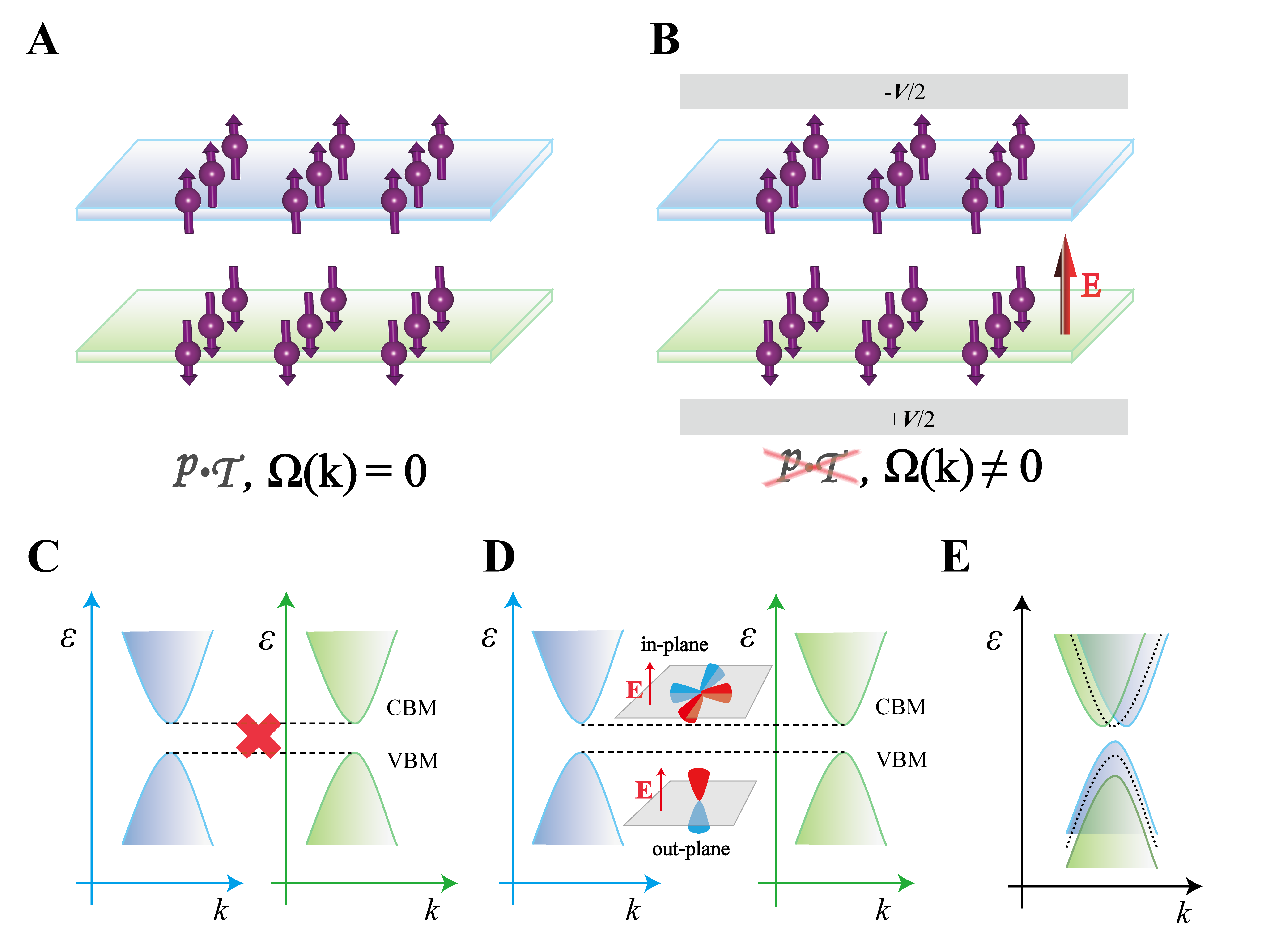}
	\caption{\textbf{Schematics of $\mathcal{PT}$ symmetry in AFM double layer.} (\textbf{A}) Schematic of AFM double layer with $\mathcal{PT}$ symmetry. The purple balls in each layer represent magnetic atoms whose magnetic orientations (out-of-plane) are depicted by arrows (e.g. Mn atoms in double-SLs MnBi$_2$Te$_4$). With $\mathcal{PT}$ symmetry, the Berry curvature $\mathbf{\Omega}(\mathbf{k})$ at each \emph{k} point is zero. (\textbf{B}) Schematic of AFM double layer with an out-of-plane electric field in which $\mathcal{PT}$ symmetry is broken. Correspondingly, $\mathbf{\Omega}(\mathbf{k})$ becomes nonzero. (\textbf{C}-\textbf{E}) Schematic band structures of AFM double-SL MnBi$_2$Te$_4$ thin film without and with electric field. In \textbf{C}, Each band is doublely degenerate and the inter-band coupling is forbidden due to $\mathcal{PT}$ symmetry. In \textbf{D} and \textbf{E}, Under the electric field, $\mathcal{PT}$ symmetry is broken and the inter-band coupling will induce a Zeeman-like splitting in the valence band and a Rashba-like splitting in the conduction band. Because the valence band maximum (VBM) and the conduction band minimum (CBM) are mainly contributed by out-of-plane and in-plane orbitals.
	 }\label{fig:1}
\end{figure}

For the bilayer AFM insulating thin film with the $\mathcal{PT}$ symmetry, such as double-SL MnBi$_2$Te$_4$, every Bloch state at any $\mathbf{k}$ point is doubly degenerated. Figure 1 demonstrates the schematics of AFM thin-film structures in real space and the corresponding electronic bands in momentum space including the influence of out-of-plane electric field. In the intrinsic film (see Figs. 1A and 1C), the Berry curvature of the $n$th Bloch band satisfies $\mathbf{\Omega}_n(\mathbf{k})=-\mathbf{\Omega}_n(\mathbf{-k})$ with $\mathcal{T}$ symmetry, and $\mathcal{P}$ symmetry enforces $\mathbf{\Omega}_n(\mathbf{k})=\mathbf{\Omega}_n(\mathbf{-k})$. Thus, $\mathbf{\Omega}_n(\mathbf{k})$ is zero in the presence of $\mathcal{PT}$ symmetry. Due to the 2D geometric property, the double-SL MnBi$_2$Te$_4$ can be applied with an out-of-plane electric field easily by using the dual-gate technology. As shown in Fig. 1B, the electrostatic potential is different in each SL, which breaks the $\mathcal{PT}$ symmetry. Therefore, the double degeneracy of each band gets broken and $\mathbf{\Omega}_n(\mathbf{k})$ becomes non-zero, which enables the possible observation of AHE via changing the carrier density in this system. In Figs. 1D and 1E, we show the schematics of the change of electronic structure under electric field. On the edge of the top valance band, the degenerated states are mainly contributed by out-of-plane orbitals with opposite spins localized at different SLs. The spacial variance of electrostatic potential can split the band edge, giving rise to a Zeeman-like band splitting. On the other hand, in-plane orbitals are major components on the edge of the bottom conduction band, and the applied electric field results in a Rashba-like spin splitting. When increasing the magnitude of electric field, the band gap decreases gradually. At the critical value, the band gap closes, which possibly leads to a topological phase transition.

By using density functional theory (DFT) (see calculation details in Methods), we calculate electronic structures of the double-SL MnBi$_2$Te$_4$ thin film with and without electric field to verify the above physical picture. Figure 2A displays the lattice structure and the first Brillouin zone (BZ) of double-SL MnBi$_2$Te$_4$. The calculated band structures and Berry curvature $\mathbf{\Omega}(\textbf{k})$ for occupied bands are shown in Figs. 2B-2K. In the absence of external field, the AFM ground state has neither $\mathcal{P}$ symmetry nor $\mathcal{T}$ symmetry, but has the $\mathcal{PT}$ symmetry and $\mathcal{C}_{3z}$ rotational symmetry along the $z$ direction. The intrinsic double-SL MnBi$_2$Te$_4$ is a trivial semiconductor with a direct band gap at the $\Gamma$ point (the calculated gap size is 76 meV) and $\mathbf{\Omega}(\textbf{k})$ is zero. We find that each SL in MnBi$_2$Te$_4$ thin film is antiferromagnetically coupled with the other in the ground state, consistent with previous calculations \cite{liJH2018,OtrokovPRL2019}.

When applying an out-of-plane electric field to the thin film, its magnetism keeps the AFM order as the ground state in a finite field range (see Fig. S2 in Supplementary Materials (SM)). But the electronic structure varies considerably. Even under a small electric field (see Fig. 2C), a sizable Zeeman-like splitting is observed at the valance bands, whose magnitude at the $\Gamma$ point is comparable to the electrostatic potential difference between adjacent SLs. And these states possess opposite spin textures. On the edge of conduction bands, the Rashba-like splitting can be observed but the splitting is much smaller compared with that of the valance bands. To understand such phenomenon, we calculate the charge density distribution in real space for split bands on the conduction and valance bands (see Fig. S3 in SM). We found that the valance band edge is mainly contributed by the $p_z$ orbital of Te atoms localized on different SLs. They are partner states under $\mathcal{PT}$ symmetry when the electric field is zero, whose spin-polarizations are locked with the SL index. The out-of-plane external field can shift these spin-polarized bands easily, and the energy splitting has the same order of magnitude as the electrostatic potential difference between different SLs. Correspondingly, these bands carry non-zero Berry curvatures. Similar argument has been used to understand the tuning of Berry curvature and valley magnetic moments in bilayer MoS$_2$ \cite{wu2013electrical}. The band edges of conduction bands are contributed by $p$ orbitals of Bi atoms and Te atoms in the same SL. The out-of-plane electric field will enhance the Rashba-like splitting, which is consistent with our DFT calculations.

\begin{figure*}
\includegraphics[width=1.5\columnwidth]{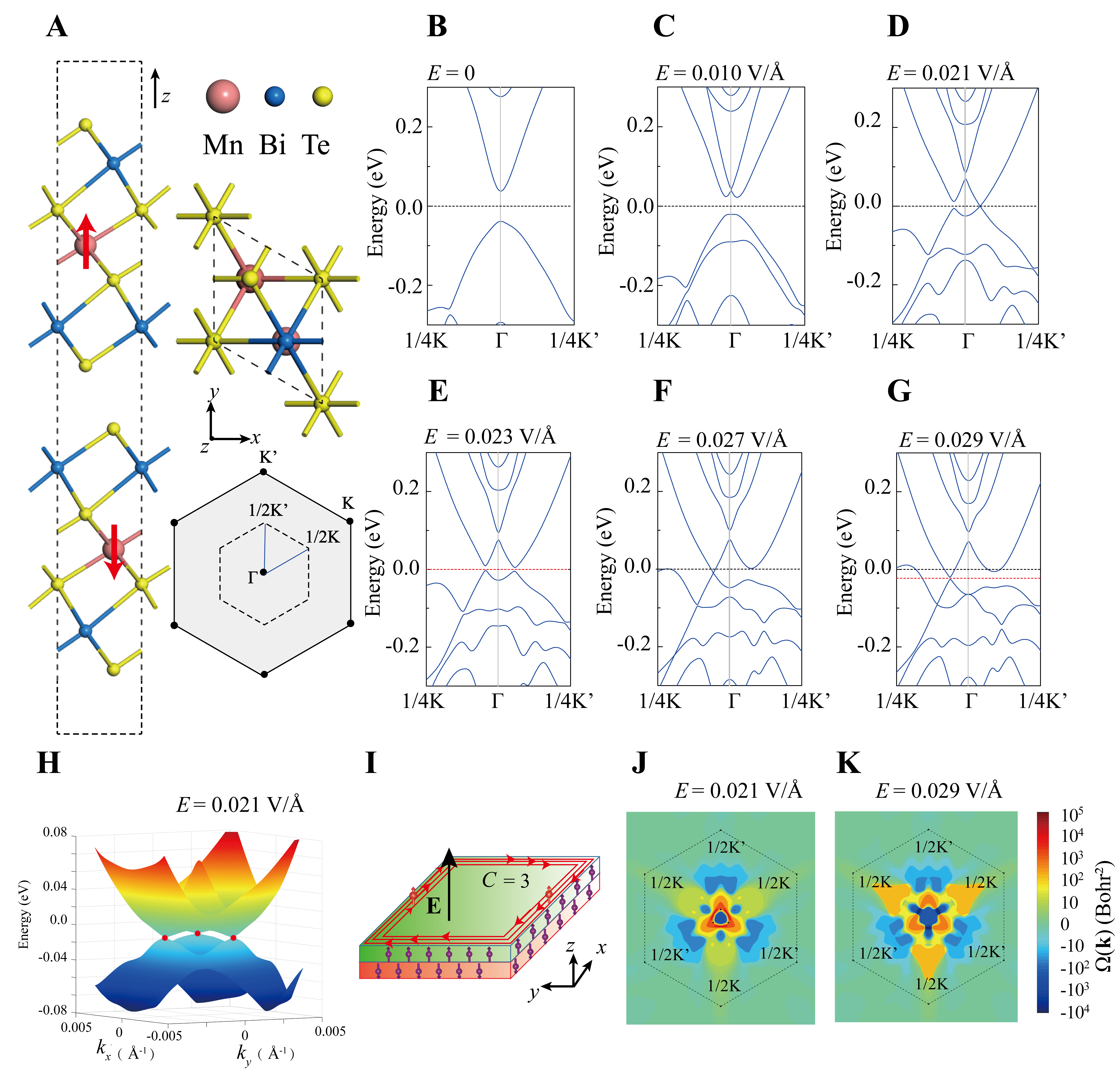}
	\caption{\textbf{Electronic band structures and Berry curvature of the double-SL MnBi$_2$Te$_4$ thin film under electric fields.} (\textbf{A}) The lattice structure of double-SL MnBi$_2$Te$_4$ thin film and the first BZ. The magnetic orientations on Mn atoms are marked by red arrows. (\textbf{B-G}) Band structures of double-SL MnBi$_2$Te$_4$ thin film under different electric fields, whose magnitudes are $0$, $0.010$, $0.021$, $0.023$, $0.027$, $0.029$ V/${\rm \AA}$, respectively. The Fermi levels marked by the black dash lines are set to be zero. It notes that in (\textbf{E}) the black dash line is overlapped with the red one. (\textbf{H}) 2D electronic structures around the Fermi level and the $\Gamma$ point when AFM thin film under the critical electric field. The band crossing points are marked as red stars. (\textbf{I}) A schematic drawing depicting the QAH edge states with a Chern number of 3 in the double-SL MnBi$_2$Te$_4$ thin film under electric field. The magnetic Mn atoms is indicated by the purple balls with arrows and the the electric field direction is indicated by black arrow. (\textbf{J}-\textbf{K}) Distributions of the Berry curvature in the first BZ for double-SLs MnBi$_2$Te$_4$ thin film under $E=0.023$ V/${\rm \AA}$ and $E=0.029$ V/${\rm \AA}$. Corresponding energy levels are marked as the red dash lines in (\textbf{E}) and (\textbf{G}).}\label{fig:2}
\end{figure*}

When increasing the electric field, band splittings of valence and conduction bands become larger and larger. We then find topological phase transitions in the double-SL MnBi$_2$Te$_4$ thin film. The first critical value $E_{c1}$ is 0.021 V/${\mathrm{\AA}}$. Figure 2D shows its band structure along high symmetric lines and the 2D electronic structure for two low-energy bands around the $\Gamma$ point is displayed in Fig. 2H. The band crossing points are along lines of $\Gamma$-K'. Because the applied electric field is along the \emph{z} direction, it does not break $\mathcal{C}_{3z}$ symmetry, then three gapless points are observed in the first BZ. Figure 2J shows the momentum resolved distribution of the Berry curvature $\mathbf{\Omega}(\textbf{k})$. Different from conventional topological phase transition that just hosts singularities of Berry curvature at the crossing points \cite{murakami2007phase}, we find a triangle region around the $\Gamma$ point with small gap size, in which all states have large $\mathbf{\Omega}(\textbf{k})$. Beyond $E_{c1}$, double-SL MnBi$_2$Te$_4$ becomes a AFM QAH system with a high Chern number of 3, as shown in Fig. 2I. To the best of our knowledge, this result is the first proposal to realize the QAH effect with high Chern numbers based on realistic AFM materials. If we further increase the electric field beyond the second critical point ($E_{c2}$ = 0.027 V/${\mathrm{\AA}}$), the gap closing process occurs along lines of $\Gamma$-K. Then the double-SL MnBi$_2$Te$_4$ becomes a trivial AFM metal. Due to the breaking of $\mathcal{PT}$ symmetry, this metallic AFM thin film has the non-zero Berry curvature distribution (Fig. 2K), contributing to an intrinsic AH signal. But its value is not quantized anymore. When we flip the direction of electric field and keep the magnetic structures, the AH signals change sign correspondingly.

\begin{figure}
\includegraphics[width=1.0\columnwidth]{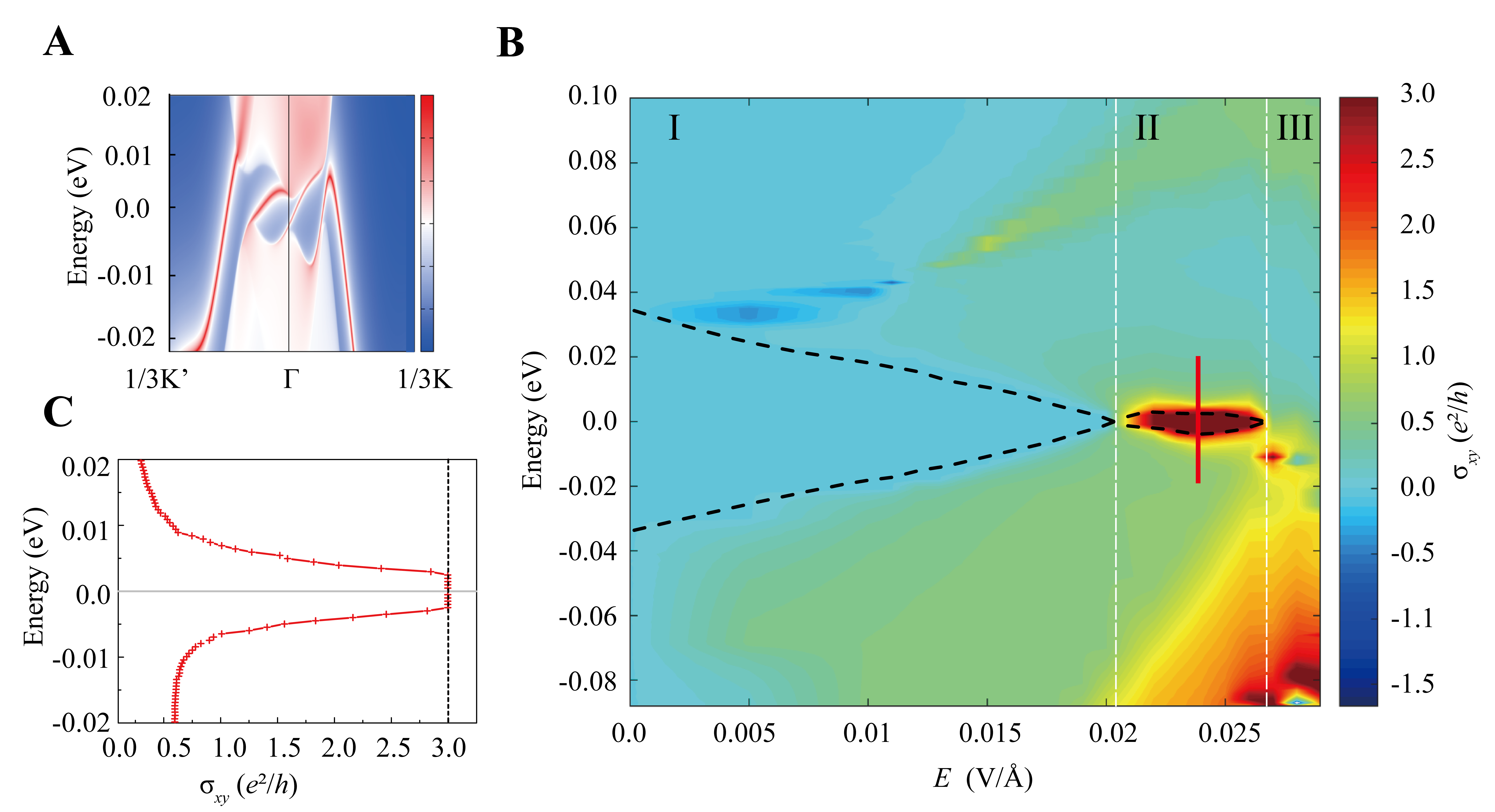}
	\caption{\textbf{Nontrivial topological properties of the double-SL MnBi$_2$Te$_4$ thin film under electric fields.} (\textbf{A}) Edge states in double-SL MnBi$_2$Te$_4$ thin film under the electric field of $E=0.023$ V/${\rm \AA}$. (\textbf{B}) The AH signal as a function of chemical potential and external perpendicular electric fields. The chemical potentials are aligned to Fermi levels. The phase regions I, II and III represent trivial semiconducting phase, QAH phase, and trivial AFM metal phase, respectively. The black dashed lines mark the energy positions of conduction and valence band edges. (\textbf{C}) AH conductance with varying chemical potential in double-SLs MnBi$_2$Te$_4$ thin film under the electric field of $E=0.023$ V/${\rm \AA}$. The chemical potential scale is marked as the red solid line in (\textbf{B}).}\label{fig:3}
\end{figure}

In order to confirm the predicted non-trivial band topology of the double-SL MnBi$_2$Te$_4$ thin film under certain electric fields, we build the 2D tight-binding model with semi-infinite boundary conditions to calculate the edge states, whose effective hopping terms are obtained from $\emph{ab initio}$ calculations. In Figs. 3A and 3C, we plot the band structure of edge states and AH conductance as a function of chemical potential in double-SL MnBi$_2$Te$_4$ thin film under the electric field of 0.023 V/${\mathrm{\AA}}$. Inside the 2D bulk gap, we can observe three chiral edge states clearly. They connect the conduction and valance bands, contributing to the quantized AH conductance of $\sigma_{xy}=3~e^2/h$ (see Fig. 3C). Beside the topologically non-trivial edge states, some other edge states are also found in Fig. 3A along the line of $\Gamma$-K. These edge states are topologically trivial without the intrinsic contribution to the AH conductance.

Figure 3B demonstrates the AH signal as a function of chemical potential and external perpendicular electric field that can be well controlled via dual-gate technology. In the trivial semiconducting phase (marked as Phase I in Fig. 3B), we observe the negative AH effect when tuning the Fermi level to the conduction band with minor electron doping. Its band structure, the Fermi surface, and related spin textures are shown in Fig. S5. We find the Rashba-like splitting on the conduction band, but different from the traditional Rashba splitting of the non-magnetic semiconductor interface, a finite band gap exists at the $\Gamma$ point without the Kramers' degeneracy. Such anti-crossing point originates from the breaking of $\mathcal{T}$ symmetry and contributes a large value to the Berry curvature, whose sign is determined by the local spin texture. Away from the anti-crossing point, the out-of-plane component in spin polarization will change the sign, resulting in the positive Berry curvature whose value is smaller compared with that at the anti-crossing point (see Fig. S5E in SM). Therefore, the total AH signal is negative. With increasing the electric field, the anti-crossing gap becomes larger and the value of corresponding negative Berry curvature becomes smaller, thus the AH signal decreases and even changes the sign. Further increasing the electric field, we obtain the quantized AH signal inside the band gap that is guaranteed by the non-trivial band topology of double-SL MnBi$_2$Te$_4$ thin film (see Phase II in Fig. 3B). The thin film becomes a trivial AFM metal in Phase III when the electric field is beyond $E_{c2}$. Its AH signal has finite value when the chemical potential is zero but increases rapidly when we shift down the Fermi level to lower energies. The corresponding band structures and Fermi surfaces are shown in Fig. S6. These states are spin-polarized and host large Berry curvature.

The drastic change of AH signal caused by varying chemical potential and electric field offers an opportunity to design AFM spintronic devices via dual-gate technology. Figures 4A and 4B show the proposed device-prototype of the AFM memory. We use the standard dual-gate to simultaneously control the perpendicular electric field and Fermi level to encode the information as the ``write-in" process and use the Hall bar to detect the AH conductance as the ``read-out" process. In principle, the AH conductance in ``Off" state is zero and its value in ``On" state could be as large as $3~e^2/h$, so its ideal on/off ratio is infinite. In the realistic experimental conditions with dissipation and at finite temperatures, the double-SL MnBi$_2$Te$_4$ thin film could be supported on substrates with large dielectric constant, such as BN, silicon, and SrTiO$_3$. The contacting interface might induce an effective electric field that slightly breaks the $\mathcal{PT}$ symmetry. The AH conductance in ``Off" state thus gains a finite value. In order to simulate such effect, we put double-SL MnBi$_2$Te$_4$ thin film on BN substrate and estimate the effective electric field strength (see Fig. S7 in SM). Then we evaluate its performance at different working temperatures. The calculated results are shown in Figs. 4C and 4D. With increasing the temperature, more states are thermally excited and the on/off ratio becomes smaller, but its value ($10^6$-$10^{14}$) still much larger than the currently realized AFM random-access memory \cite{Zelezny2014,ZhouXF2018,marti2014room,Moriyama2018,wadley2016electrical,bodnar2018writing,olejnik2017}. In our current calculations, we mainly consider the intrinsic Berry-phase-related AHE. Some other extrinsic mechanisms (e.g. skew scattering) may also contribute to AH signals \cite{Nagaosa2010}, but their contributions should be insignificant. Because, in contrast to the magnetic alloys, MnBi$_2$Te$_4$ thin films are cleaved from high-quality single crystals with lower density of impurities and disorders \cite{deng2019magnetic,liu2019quantum,ge2019}. The intrinsic mechanism should dominate the contribution of AHE.

\begin{figure}
\includegraphics[width=1.0\columnwidth]{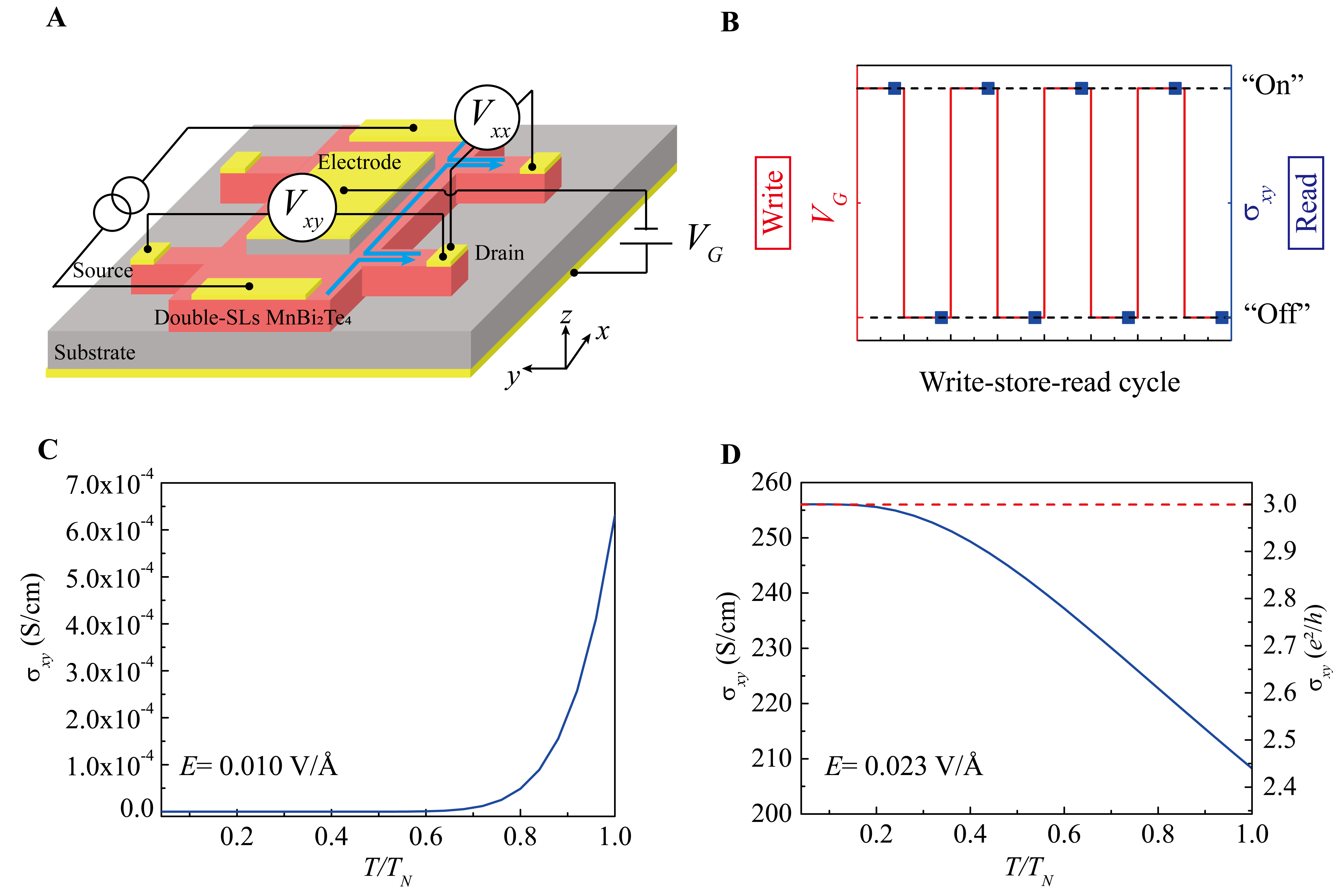}
	\caption{\textbf{Memory devices based on AFM double-SL MnBi$_2$Te$_4$ thin film.} (\textbf{A}) Top view of schematics for a Hall bar device. The blue arrows indicate the edge current direction. The red areas show the double-SL MnBi$_2$Te$_4$ thin film. The gray and yellow areas represent the substrate and attached electrodes, respectively. (\textbf{B}) The schematic operation for the AFM memory device. ``On" and ``Off" states are for double-SL MnBi$_2$Te$_4$ thin film in the trivial semiconductor phase and QAH phase. (\textbf{C}-\textbf{D}) AH conductance with varying temperature in double-SL MnBi$_2$Te$_4$ thin film under the electric field of $E=0.010$ V/${\rm \AA}$ and $E=0.023$ V/${\rm \AA}$, the temperature is normalized to N\'eel temperature $T_N$=25K. }\label{fig:4}
\end{figure}

Beyond the double-SL, the discussed physics and proposed spintronic device-prototype in this work could be generalized to other MnBi$_2$Te$_4$ thin films with even-SL thickness as long as $\mathcal{PT}$ symmetry is not intrinsically broken. As the confirmation, we apply the external electric field to a four-SL MnBi$_2$Te$_4$ thin film that was regarded as an axion insulator \cite{zhang2018topological}. The calculated electronic structures are shown in Fig. S8. We can still observe the spin-polarized band splitting and topological phase transitions driven by the electric field. While, its critical values are smaller than those in double-SL MnBi$_2$Te$_4$ and the topologically non-trivial phase region becomes narrower. To conclude, we believe that the electric field is an efficient method to manipulate the Berry curvature effects in even-SL MnBi$_2$Te$_4$ thin films, resulting in a large change of AH signal. Thus the even-SL MnBi$_2$Te$_4$ thin film is a promising material platform to build the low-power AFM memory bit based on the AH signal with electric write-in and read-out. We expect its performance to be stable and robust under the external magnetic field. This work paves the way for using even-SL MnBi$_2$Te$_4$ thin films, and perhaps AFM topological thin films more generally, in a new generation of electrically switchable AFM spintronic devices.

S.D., J.L., Z.L., Y.X., and W.D. acknowledge financial supports from the Basic Science Center Project of NSFC (Grant No. 51788104), the Ministry of Science and Technology of China (Grants No. 2016YFA0301001, No. 2018YFA0307100, and No. 2018YFA0305603), the National Natural Science Foundation of China (Grants No. 11674188 and No. 11874035), and the Beijing Advanced Innovation Center for Future Chip (ICFC). A.R. and P.T. acknowledge financial supports from the European Research Council (ERC-2015-AdG-694097). P.T. acknowledges the received funding from the European Union Horizon 2020 research and innovation programme under the Marie Sklodowska-Curie grant agreement No 793609. The Flatiron Institute is a division of the Simons Foundation.

\end{document}